Running head: ONTOLOGY-BASED ANNOTATION OF MULTIMEDIA LANGUAGE DATA FOR THE SEMANTIC WEB

# Ontology-Based Annotation of Multimedia Language Data for the Semantic Web

Artem Chebotko, Shiyong Lu and Farshad Fotouhi

Wayne State University

Department of Computer Science

5143 Cass Avenue, Detroit, Michigan 48202, USA

{artem, shiyong, fotouhi}@ wayne.edu

Anthony Aristar

Wayne State University

Program in Linguistics

51 W.Warren, Detroit, Michigan 48202, USA

aristar@linguistlist.org

## Inside Chapter

There is an increasing interest and effort in preserving and documenting endangered languages. Language data are valuable only when they are well-cataloged, indexed and searchable. Many language data, particularly those of lesser-spoken languages, are collected as audio and video recordings. While multimedia data provide more channels and dimensions to describe a language's function, and gives a better presentation of the cultural system associated with the language of that community, they are not text-based or structured (in binary format), and their semantics is implicit in their content. The content is thus easy for a human being to understand, but difficult for computers to interpret. Hence, there is a great need for a powerful and user-friendly system to annotate multimedia data with text-based, well-structured and searchable metadata. This chapter describes an ontology-based multimedia annotation tool, *OntoELAN*, that enables annotation of language multimedia data with a linguistic ontology.

## Introduction

Recently, there is an increasing interest and effort for preserving and documenting endangered languages (Lu et al., 2004; The National Science Foundation, 2004). Many languages are in serious danger of being lost and if nothing is done to prevent it, half of the world's approximately 6,500 languages will disappear in the next 100 years. Most languages will disappear without a trace, along with the cultural and scientific information they contain, unless we act immediately to collect, analyze and archive language documentation.

When a language disappears, there are two losses. First, there is the loss of valuable scientific data about the cultural system that produced the language. The death of a language symbolizes the passing away of a community's traditional poetry, songs, images, stories, cultural traditions and religious rites. Second, any language loss represents a serious scientific loss: studies of linguistic diversity and cross-linguistic comparisons drive much of linguistic theory. In addition, linguistic material provides valuable information about population movements, contacts, and genetic relationships. Thus, in the face of the pressing threat to so many of the world's languages, we must not only preserve existing language documentation but also encourage the collection of more. Furthermore, we must make this documentation easily accessible, since in order to derive maximum scientific benefit from the information, the documentation must be shared among different research communities.

Language data are valuable only when they are well-cataloged, indexed, and searchable. Many language data, particularly those of lesser-spoken languages, are collected as audio and video recordings. While multimedia data provide more channels and dimensions to describe a language's function, and give a better presentation of the cultural system associated with the language of that community, they are not text-based or structured (in binary format), and their semantics is implicit in their content. The content is thus easy for a human being to understand, but difficult for computers to interpret. Hence, there is a great need for a powerful and user-friendly system to annotate multimedia data with text-based, well-structured and searchable metadata. However, different annotators might use different vocabularies to annotate multimedia data, which causes low recall and precision in further search and retrieval. We propose an ontology-based annotation approach, in which a linguistic ontology is used so that the terms and their relationships are formally defined. In this way, annotators will use the same vocabulary to annotate data, enabling ontology-driven search engines to retrieve multimedia data with greater recall and precision. We believe that, even though in a particular domain, it can be very difficult to enforce a uniform ontology that is agreed on by the whole community, ontology-based annotation will benefit the community once ontology-aware federated retrieval systems are developed based on techniques such as ontology mapping, alignment and merging (Klein, 2001).

We present an ontology-based linguistic multimedia annotation tool, *OntoELAN* – a successor of EUDICO Linguistic Annotator (*ELAN*) (Hellwig & Uytvanck, 2004), that was developed at the Max Planck Institute for Psycholinguistics, Nijmegen, Netherlands, with the aim to provide a sound technological basis for the annotation and exploitation of multimedia recordings. Although *ELAN* is designed specifically for the linguistic domain (analysis of language, sign language and gesture), ELAN can be used for annotation, analysis and documentation purposes in other domains that involve multimedia data.

*OntoELAN* inherits all *ELAN*'s features and extends the tool with an ontology-based annotation functionality. In particular, our main contributions are:

- *OntoELAN* can open and display ontologies specified in Web Ontology Language (OWL) (Bechhofer et al., 2004);
- *OntoELAN* supports the creation of a language profile, which enables a user to choose a subset of terms from a linguistic ontology and conveniently rename them if needed.
- *OntoELAN* supports the creation of an ontological tier, which can be annotated with profile terms and, therefore, their corresponding ontological terms.
- *OntoELAN* saves annotations in XML (Bray, Paoli, Sperberg-McQueen, Maler & Yergeau, 2004) format as instances of classes from the General Multimedia Ontology, which is designed by us based on the XML schema (Fallside, 2001) for *ELAN* annotation files.

- *OntoELAN*, while annotating an ontological tier, creates instances of classes from the ontology associated with the ontological tier and relates them to instances of classes from the General Multimedia Ontology.

Since *OntoELAN* is developed to fulfill annotation requirements for the linguistic domain, it is natural for us to use linguistic annotation examples and link the General Ontology for Linguistic Description (GOLD) to an ontological tier. To the best of our knowledge, *OntoELAN* is the first audio/video annotation tool in the linguistic domain that supports ontology-based annotation. It is expected that the availability of such a tool will greatly facilitate the creation of linguistic multimedia repositories as islands of the Semantic Web of language engineering.

## Related Work

In the following, first, we identify the requirements for linguistic multimedia annotation. Then, we review existing annotation tools with respect to these requirements. Finally, we conclude that the tools do not fully satisfy our requirements, and this motivates our development of *OntoELAN*.

The linguistic domain places some minimum requirements on multimedia annotation tools. While semantics-based content, such as speeches, gestures, signs and scenes, are important, color and shape are less important. To annotate semantics-based content, a tool should provide a time axis and the capability of its subdivision into time slots/segments, multiple tiers for different semantic content. Obviously, there should be some multimedia resource metadata, such as title, authors, date and time. Additionally, a tool should provide ontology-based annotation features to enable the same annotation vocabulary for a particular domain.

As related work, we give a brief review of the following tools: *Protégé* (Stanford University, 2004), *IBM MPEG-7 Annotation Tool* (International Business Machines Corporation, 2004) and *ELAN* (Hellwig & Uytvanck, 2004).

*Protégé* is a popular ontology construction and annotation tool that is developed at Stanford University. *Protégé* supports the Web Ontology Language through an OWL Plugin, which allows a user to load an OWL ontology, annotate data and save annotation markup. Unfortunately, *Protégé* provides only simple multimedia support through the Media Slot Widget. The Media Slot Widget allows the inclusion and display of video and audio files in *Protégé*, which may be enough for general description of multimedia files like metadata entries, but not sufficient for annotation of a speech, where the multimedia time axis and its subdivision into segments are crucial.

*IBM MPEG-7 Annotation Tool* was developed by IBM to assist the annotation of video sequences using MPEG-7 (Martínez, 2003) metadata based on the shots of the video. It does not support any ontology language, and uses an editable lexicon from which a user can choose keywords to annotate shots. A *shot* is defined as a time period in video in which the frames have similar scenes. Annotations are saved based on MPEG-7 XML Schema (Martínez, 2003). However, this shot and lexicon based annotation approach does not provide enough flexibility for linguistic multimedia annotation. In particular, the shot approach is good for the annotation of content-based features like color and texture, but not for time alignment and time segmentation required for semantics-based content annotation.

*ELAN* (EUDICO Linguistic Annotator) developed at the Max Planck Institute for Psycholinguistics, Nijmegen, Netherlands, is designed specifically for the analysis of spoken language, sign language and gesture to provide a sound technological basis for the annotation and exploitation of multimedia recordings. *ELAN* provides many important features for linguistic data annotation, such as time segmentation and multiple annotation layers, but not the support of ontology-based annotation. Annotation files are saved in the XML format based on *ELAN*'s XML schema.

As a summary, existing annotation tools, such as *Protégé* and *IBM MPEG-7 Annotation Tool*, are not suitable for our purpose since they do not support many multimedia annotation operations such as multiple tiers, time transcription and translation of linguistic audio and video data. *ELAN* is the best candidate as a linguistic multimedia annotator, and it is already used by linguists throughout the world. *ELAN* provides most of the required features for linguistic multimedia annotation, which motivates us to use it as the basis for the development of *OntoELAN* to add ontology-based annotation features, such as the support of an ontology and a language profile.

## An Overview of OntoELAN

*OntoELAN* is an ontology-based linguistic multimedia annotator developed on the top of the *ELAN* annotator. It was partially sponsored and developed as a part of the Electronic Metastructure for Endangered Languages Data (E-MELD) project. Currently, the *OntoELAN* source code contains more than 60,000 lines of Java code and has several years of the development history started by the Max Planck Institute for Psycholinguistics team and continued by the Wayne State University team. Both development teams will continue their collaboration on *ELAN* and *OntoELAN*.

*OntoELAN* has the long list of technical features including the following features that are inherited from *ELAN*:

- display a speech and/or video signals, together with their annotations;
- time linking of annotations to media streams;
- linking of annotations to other annotations;
- unlimited number of annotation tiers as defined by a user;
- different character sets;
- basic search facilities.

*OntoELAN* implements the following additional features:

- loading of OWL ontologies;
- language profile creation;
- ontology-based annotation;
- storing annotations in the XML format based on the General Multimedia Ontology and domain ontologies.

The main window of *OntoELAN* is shown in Figure 1. *OntoELAN* has the video viewer, the annotation density viewer, the waveform viewer, the grid viewer, the subtitle viewer, the text viewer, the timeline viewer, the interlinear viewer and associated with them controls and menus. All viewers are synchronized so that whenever a user accesses a point in time in one viewer, all the other viewers move to the corresponding point in time automatically. The video viewer displays video in “mpg” and “mov” formats and can be resized or detached to play video in a separate window. The annotation density viewer is useful for navigation through the media file and analysis of annotations concentration. The waveform viewer displays the waveform of the audio file in “wav” format; in case of video files, there should be an additional “wav” file present to display waveform. The grid viewer displays annotations and associated time segments for a selected annotation tier. The subtitle viewer displays annotations on selected annotation tiers at the current point in time. The text viewer displays annotations of a selected annotation tier as a continuous text. The timeline viewer and the interlinear viewer are interchangeable and both display all tiers and all their annotations; only one viewer can be used at a time. In this paper, we will mostly work with the timeline viewer (see Figure 1), which supports the operations on tiers and annotations. Because a significant part of *OntoELAN* interface is inherited from *ELAN*, the reader can refer to Hellwig and Uytvanck (2004) for a detailed description.

*OntoELAN* uses and manages several data sources:

- General Multimedia Ontology (OWL) – ontological terms for multimedia annotations.
- Linguistic domain ontologies (OWL) – ontological terms for linguistic annotations.
- Language profiles (XML) – a selected subset of domain ontology terms for linguistic annotations.
- *OntoELAN* annotation documents (XML) – storage for linguistic multimedia annotations.

The data flow diagram for *OntoELAN* is shown in Figure 2. We do not specify the names of most data flows because they are too general to give any additional information. Two data flows from a user are user-defined terms for language profiles and linguistic multimedia annotations.

In the following sections, we will present more details on *OntoELAN* data sources and data flows. We focus more on the description of features that make *OntoELAN* an ontology-based multimedia annotator, like OWL support, a linguistic domain ontology, the General Multimedia Ontology, a language profile, ontological annotation tiers, and so forth.

## Support of OWL

OWL Web Ontology Language (Bechhofer et al., 2004) is recently recommended as a semantic markup language for publishing and sharing ontologies on the World Wide Web. It is developed as a revision of DAML+OIL language and has more expressive power than XML, RDF and RDF Schema (RDF-S). OWL provides constructs to define ontologies, classes, properties, individuals, data types and their relationships. In the following, we present a brief overview of the OWL constructs and refer the reader to (Bechhofer et al., 2004) for more details.

*Classes*. A class defines a group of individuals that share some properties. A class is defined by *owl:Class* and different classes can be related by *rdfs:subClassOf* into a class hierarchy. Other relationships between classes can be specified by *owl:equivalentClass*, *owl:disjointWith*, and so forth. The extension of a class can be specified by *owl:oneOf* with a list of class members or by *owl:intersectionOf*, *owl:unionOf* and *owl:complementOf* with a list of other classes.

*Properties*. A property states relationships between individuals or from individuals to data values. The former is called *ObjectProperty* and specified by *owl:ObjectProperty*. The latter is called *DatatypeProperty* and specified by *owl:DatatypeProperty*. Similarly to classes, different properties can be related by *rdfs:subPropertyOf* into a property hierarchy. The domain and range of a property are specified by *rdfs:domain* and *rdfs:range*, respectively. Two properties might be asserted to be equivalent by *owl:equivalentProperty*. In addition, different characteristics of a property can be specified by *owl:FunctionalProperty*, *owl:InverseFunctionalProperty*, *owl:TransitiveProperty*, and *owl:SymmetricProperty*.

*Property restrictions*. A property restriction is a special kind of a class description. It defines an anonymous class, namely the set of individuals that satisfy the restriction. There are two kinds of property restrictions: *value constraints* and *cardinality constraints*. Value constraints restrict the values that a property can take within a particular class, and they are specified by *owl:allValuesFrom*, *owl:someValuesFrom*, and *owl:hasValue*. Cardinality constraints restrict the number of values that a property can take within a particular class, and they are specified by *owl:minCardinality*, *owl:maxCardinality* and *owl:cardinality*.

OWL has three species (in increasingly-expressive order): OWL Lite, OWL DL and OWL Full. OWL Lite places some limitations on the usage of constructs and is primarily suitable for expressing taxonomies. For example, *owl:unionOf* and *owl:complementOf* are not part of OWL Lite and cardinality constraints may only have 0 or 1 value. OWL DL provides more expressivity and still guarantees computational completeness and decidability. In particular, OWL DL supports all OWL constructs, but places some additional restrictions, e.g., a class cannot be treated as an individual. Finally, OWL Full gives the maximum expressiveness, but does not guarantee computational tractability.

*OntoELAN* uses the Jena 2 Java framework (Hewlett-Packard Labs, 2004) for writing Semantic Web applications to provide OWL DL support. On a language profile creation stage, *OntoELAN* uses the class hierarchy of an ontology specified with *rdfs:subClassOf* constructs to display the ontology. On a data annotation stage, *OntoELAN* handles the semantics of other OWL constructs to provide a dynamic user interface for creating instances of ontology classes, assigning property values, and so forth.

## Linguistic Domain Ontology

We use the General Ontology for Linguistic Description (GOLD) (Farrar & Langendoen, 2003) as an example of a linguistic domain ontology. To make things clear from the beginning, *OntoELAN* does not require GOLD itself, but loads and supports any other linguistic domain ontology at runtime. Thus *OntoELAN* can be used as a multimedia annotator in other domains that require similar annotation features. Moreover, a user can load different ontologies for distinct annotation tiers to enable annotation with terms that come from multiple ontologies and even multiple domains. For example, a gesture ontology can be used for linguistic multimedia annotation as speaker's gestures help to understand the meaning of a speech. Therefore, linguists can use GOLD in one tier and the gesture ontology in another tier to capture more semantics.

The General Ontology for Linguistic Description is an ongoing research effort lead by the University of Arizona to define linguistic domain specific terms using OWL. GOLD is constantly under revision: not only new classes and properties are introduced, but also the relations between existing classes/properties may change. Current information about GOLD is available at http://www.emeld.org/ and the ontology is also downloadable from http://www.u.arizona.edu/~farrar/gold.owl. We briefly describe the GOLD content in the next few paragraphs and refer the reader to Farrar and Langendoen (2003) and also to Farrar (2004) for more details.

GOLD provides a semantic framework for the representation of linguistic knowledge and organizes knowledge into four major categories:

- *Expressions* – physically accessible aspects of a language. Linguistic expressions include the actual printed words or sounds produced when someone speaks. For example, *OrthographicExpression*, *Utterance*, *SignedExpression*, *Word*, *WordPart*, *Prefix*.
- *Grammar* – the abstract properties and relations of a language. For example, *Tense*, *Number*, *Agreement*, *PartOfSpeech*.
- *Data structures* – constructs that are used by linguists to analyze language data. A linguistic data structure can be viewed as a structuring mechanism for linguistic data content. For example, a lexical entry is a data structure used to organize lexical content. Other examples are a phoneme table and a syntactic tree.

- *Metaconcepts* – the most basic concepts of linguistic analysis. The example of a metaconcept is a language itself.

In our examples, we only use simple concepts found in GOLD, such as *Noun*, *Verb*, *Participle*, *Preverb*. These terms are the subclasses of *PartOfSpeech* and their meanings are easy to understand without further explanation. Additionally, we use concepts *Animate* (living things, including humans, animals, spirits, trees, and most plants) and *Inanimate* (non-living things, such as objects of manufacture and natural "non-living" things), which are two grammatical genders or classes of nouns.

## General Multimedia Ontology

Although *OntoELAN* is an ontology-based annotator, nothing prevents a user not to use ontological terms for annotation. In fact, for linguistic multimedia annotation, there should usually be several annotation tiers whose annotations are not based on ontological terms. For example, a speech transcription and a speech translation into another language do not use any ontology. Consequently, *OntoELAN* needs to save not only instances of classes created for ontology-based annotations, but also text-based data created without ontologies. One solution is to use XML Schema to save an annotation file in the XML format – this approach is exploited by *ELAN*. To fully conform the ontology-based annotation approach, we provide different solution – a multimedia ontology.

We have developed the multimedia ontology that we called General Multimedia Ontology and that serves as a semantic framework for multimedia annotation. In contrast to domain ontologies, the General Multimedia Ontology is a crucial component of the system. *OntoELAN* saves its annotations in the XML format as instances of classes found in the General Multimedia Ontology and linguistic domain ontologies that are used in ontological tiers.

The General Multimedia Ontology is specified in Web Ontology Language and is designed based on *ELAN*'s XML schema for annotation. The General Multimedia Ontology contains the following classes:

- *AnnotationDocument*, which represents the whole annotation document.
- *Tier*, which represents a single annotation tier/layer. There are several predefined types of tiers that a user can choose.
- *TimeSlot*, which represents the notion of a time segment that is contained in a tier.
- *Annotation*, which represents an annotation and can be either *AlignableAnnotation* or *ReferringAnnotation*.
- *AlignableAnnotation*, which links directly to a time slot.
- *ReferringAnnotation*, which can reference an existing *AlignableAnnotation*.
- *AnnotationValue*, which has two subclasses: *StringAnnotation* and *OntologyAnnotation* that represent two different ways of annotating.
- *MediaDescriptor*, *TimeUnit* and others.

Relationships among major classes found in the General Multimedia Ontology are presented in Figure 3. In general, *AnnotationDocument* may have zero or many *Tiers*, which, in turn, may have zero or many *Annotations*. *Annotation* can be either *AlignableAnnotation* or *ReferringAnnotation*, where *AlignableAnnotation* can be divided by *TimeSlots*, and *ReferringAnnotation* can refer to another annotation. *ReferringAnnotation* may refer to *AlignableAnnotation* as well as to *ReferringAnnotation*, but the root of the referenced annotations must be an *AlignableAnnotation*. Each *Annotation* has one *AnnotationValue,* which can be either *StringAnnotation* or *OntologyAnnotation*. *StringAnnotation* represents any string which a user can input as an annotation value, but values, represented by *OntologyAnnotation*, come from a language profile and, consequently, from an ontology. Note that the General Multimedia Ontology allows *OntologyAnnotation* to be used only with *ReferringAnnotation*. In other words, tiers with *AlignableAnnotations* do not support ontology-based annotation. This limitation is due to software development issues – *OntoELAN* does not support annotation with ontological terms in alignable tiers. We intentionally emphasize this constraint in the ontology, although conceptually it should not be the case.

Among our contributions is the introduction of the *OntologyAnnotation* class, which serves as an annotation unit for an ontology-based annotation. *OntologyAnnotation* has restrictions on the following properties:

- *hasOntAnnotationId* – the ID of the annotation. The property cardinality equals one (*owl:cardinality* = *1*).
- *hasUserDefinedTerm*, which relates *OntologyAnnotation* to a term in a language profile (described in the next section). The property cardinality equals one (*owl:cardinality* = *1*).
- *hasInstances*, which relates *OntologyAnnotation* to a term (represented as an instance) in an ontology used for annotation. The property cardinality is greater than zero (*owl:minCardinality* = *1*).
- *hasOntAnnotationDescription* – descriptions/comments on the annotation. The property cardinality is not restricted.

The General Multimedia Ontology is available at http://database.cs.wayne.edu/proj/OntoELAN/multimedia.owl. We will add new concepts to the

ontology in case if *OntoELAN* needs them for annotation. We have developed the General Multimedia Ontology especially for *OntoELAN* and have not included most concepts in the multimedia domain. In particular, we did not include multimedia concepts such as those related to shapes, colors, motions, audio spectrum, etc. Our relatively small ontology focuses on high-level multimedia annotation features and can be used for similar annotation tasks.

## Language Profile

A language profile is a subset of ontological terms, possibly renamed, that are used in the annotation of a particular multimedia resource. The idea of a language profile comes from the following practical issues related to an ontology-based annotation.

A domain ontology defines all terms related to a particular domain, and the number of terms is usually considerably large. However, to annotate a concrete data resource of a particular language, an annotator usually does not need all terms from an ontology. An experienced annotator can identify a subset of ontological terms that can be useful for a given resource and a particular annotation task. For example, an annotator may only use a subset of GOLD to annotate a particular language recording, while he may need a different subset for another language.

Linguists have been annotating multimedia data without the standardized set of terms that an ontology provides. They have their individual sets of terms that they are accustomed to use for annotation. It will be difficult to come to a consensus about class names in GOLD, so that every linguist is satisfied with them. Therefore, we should allow users to rename standard terms into user-defined terms in a language profile. Additionally, linguists widely use abbreviations like “n” for “noun” which is concise and convenient. Finally, linguists whose native language is, for example, Ukrainian, may prefer to use annotation terms in Ukrainian rather than in English.

More formally, a language profile is defined as a quadruple: ontological terms; user-defined terms; a mapping between ontological terms and user-defined terms; a reference to an ontology that contains the descriptions of terms. In *OntoELAN*, a language profile provides convenience and flexibility for a user to:

- select the subset of ontological terms useful for a particular resource annotation;
- rename ontological terms, e.g., use another language, give an abbreviation or a synonym;
- combine the meaning of two or many ontological terms and assign it to one user-defined term (e.g., ontological terms “Inanimate” and “Noun” may be combined and renamed as “NI”).

*OntoELAN* supports ontology-based annotation by means of a language profile. A user opens an ontology, creates a profile and links it to an ontological tier. Annotation values for an ontological tier can only be selected from a language profile.

A language profile is stored as an XML document (see Figure 4) with a predefined schema, which contains term mappings, a link to a source ontology and some additional information about the profile and its author. A user can easily create, open, edit and save language profiles with *OntoELAN*.

Figure 4 presents an example language profile, created by the first author of this book chapter and linked to the GOLD ontology at URL http://www.u.arizona.edu/~farrar/gold.owl. In this example, there is only one user-defined term “NI” that maps to ontological terms “Noun” and “Inanimate”. This is a one-to-many mapping, but a mapping can be many-to-many as well. For example, we can add another user-defined term “IN” that maps to the same ontological terms “Noun” and “Inanimate”. In general, a mapping can be one-to-one, one-to-many, many-to-one or many-to-many.

## Annotation Tiers and Linguistic Types

*OntoELAN* allows a user to create an unlimited number of annotation tiers. Multiple tier feature is a must for linguistic multimedia annotation. For example, while annotating an audio monolog, a linguist may choose separate tiers to write a monolog transcription, a translation, a part of speech annotation, a phonetic transcription, and so forth.

An annotation tier can be either *alignable* or *referring*. Alignable tiers are directly linked to the time axis of an audio/video clip and can be divided into segments (time slots); referring tiers contain annotations that are linked to an annotation on another tier, which is called *a parent tier* and can be alignable or referring. Thus, tiers can be viewed as a hierarchy, where its root must be an alignable tier. Following the previous example, the speech transcription could be an independent time-alignable tier divided into time slots of speaker's utterances. On the other hand, the translation referring tier could refer to the transcription tier, so that the translation tier inherits its time alignment from the transcription tier.

After a tier hierarchy is established, changes in one tier may influence other tiers. Deletion of a parent tier is cascaded, such that all its child tiers are automatically deleted. This is also true about annotations on a tier: deletion of an annotation on a parent tier causes the deletion of all corresponding annotations on its child tiers. Alteration of the time slot on a parent tier influences all child tiers as well.

Each annotation tier has associated with it a linguistic type. There are five predefined linguistic types in *OntoELAN*, which put some constraints on tiers assigned to them. The predefined linguistic types, also called stereotypes, are:

- *None:* The annotation on the tier is linked directly to the time axis. This is the only type that alignable tiers can have.
- *Time Subdivision:* The annotation on the parent tier can be subdivided into smaller units, which, in turn, can be linked to time slots. They differ from annotations on alignable tiers in that they are assigned to a slot that is contained within the slot of their parent annotation.
- *Symbolic Subdivision:* Similar to the previous type, but the smaller units cannot be linked to the time slots.
- *Symbolic Association:* The annotation on the parent tier cannot be subdivided further, so there is a one-to-one correspondence between the parent annotation and its referring annotation.
- *Ontological Type:* The annotation on such a tier is linked to a language profile. This is not an independent type as it can be used only in combination with referring tier types such as *Time Subdivision*, *Symbolic Subdivision* or *Symbolic Association*. To emphasize that a referring tier allows ontology-based annotation, we call it an ontological tier.

Only ontological tiers allow annotation based on language profile terms; other types of tiers allow annotation with string values.

## Linguistic Multimedia Annotation with OntoELAN

In this section, we describe an annotation process in *OntoELAN* using a linguistic multimedia resource annotation example. In general, an annotation process in *OntoELAN* consists of three major steps: (1) language profile creation; (2) creation of tiers; and (3) creation of annotations. The first step is unnecessary if ontological tiers will not be defined. The second step can be completed partially for non-ontological tiers before the creation of a language profile. It is also possible to have multiple profiles for multiple ontological tiers, but there is always one-to-one correspondence between a profile and an ontological tier.

As an example, we annotate an audio file, which contains a sentence in Potawatomi, one of the North American native languages.

We first load the GOLD ontology and create the Potawatomi language profile. Figure 5 presents a snapshot of the profile creation window. Tabs “Index” and “Ontology Tree” on the left provide two views of an ontology: a list view, which displays all the terms in the ontology in the alphabetical order; a tree view, which displays the class hierarchy of the ontology. A user can select required terms from one of these views, add the selected terms to the “Ontological Terms” list and rename them as shown in the “User Defined Terms” list. For example, in Figure 5, we select ontological terms “Inanimate” and “Noun” and combine them under one user-defined term “NI”.

After the language profile is created, we define six tiers in the *OntoELAN* main window (see Figure 6):

- *Orthographic* of type “None” (linked to the time axis);
- *Translation* of type “Symbolic Association” (referring to *Orthographic*);
- *Words* of type “Symbolic Subdivision” (referring to *Orthographic*);
- *Parse* of type “Symbolic Subdivision” (referring to *Words*);
- *Gloss* of type “Symbolic Association” (referring to *Parse*);
- *Ontology* of type “Symbolic Association” and “Ontological Type” (referring to *Gloss*).

The created tier hierarchy is shown in Figure 7.

Finally, we specify annotation values on the six tiers (see Figure 6). We annotate the *Orthographic* tier first, because it is the root of the tier hierarchy, and its time alignment is inherited by other tiers. We do not divide the *Orthographic* tier into time slots and its time axis contains the whole sentence in Potawatomi. The *Translation* tier inherits time alignment from its parent and cannot be subdivided further (type “Symbolic Association”). The *Words* tier also inherits time alignment from *Orthographic*, but, in this case, we subdivide it into segments that correspond to words in the sentence. Similarly, we subdivide the *Parse* tier alignment inherited from *Words*. The *Gloss* tier inherits alignment from *Parse*, and the *Ontology* tier inherits alignment from *Gloss*; both *Gloss* and *Ontology* do not allow further subdivision. Correct alignment inheritance is important, because there is a semantic correspondence between segments of different tiers. For example, the Potawatomi word “neko” in the *Words* tier has the corresponding gloss “used to” in the *Gloss* tier and the part of speech “PC” (maps to the GOLD *Participle* concept) in the *Ontology* tier.

All the annotations are represented by string values, except for the annotations on the *Ontology* tier, which is defined as an ontological tier. Unlike with the string value annotations, a user annotates an ontological tier by selecting user-defined terms from a profile. Once such term is selected, the next step is creating individuals of the corresponding ontological term(s). If the ontological term is defined as a class instance in the ontology, the user is not required to perform any additional operations. Otherwise, the user must input a class instance identifier and assign values to the corresponding properties, which in turn may need the creation of other class instances.

The resulting annotation document is saved in XML format as instances of the General Multimedia Ontology classes and the GOLD classes. The example of the XML markup for the *Ontology* tier instance and the referring annotation instance with the identifier (ID) “a42” on that tier is shown in Figure 8. Several properties are defined for the *Ontology* tier, such as ID, parent tier, profile and linguistic type. For the referring annotation, ID, reference to another annotation, and annotation value are defined. The annotation value has the *OntologyAnnotation* class instance with some ID, the user-defined term “PV” and the reference to the GOLD concept *Preverb*, which is a class instance in the ontology. The markup in Figure 8 is based on the General Multimedia Ontology and has the reference to the GOLD class instance mentioned above.

## Conclusions and Future Work

In this chapter, we addressed the challenge of multimedia annotation for the Semantic Web of language engineering. Our main contribution is the development of *OntoELAN*, a linguistic multimedia annotation tool, which employs the ontology-based annotation approach. *OntoELAN* is the first attempt at annotating linguistic multimedia data with a linguistic ontology. Meanwhile, the ontological annotations share the data on the linguistic ontologies. Our future work will provide more channels for sharing data on the Web, such as multimedia descriptions, the language words, and so forth. Also, we need to improve the current search features of the tool, which include text search and retrieval in one annotation document to search, retrieve and compare the linguistic multimedia annotation data on the Web. Moreover, we plan to integrate text document annotation capabilities into *OntoELAN* and include semi-automatic annotation support, similar to *Shoebox* (SIL International, 2000).

**Acknowledgements**

We would like to thank Yu Deng, a Master student from Wayne State University, for her hard work on the *OntoELAN* implementation. We are grateful to Hennie Brugman, Alexander Klassmann, Han Sloetjes, Albert Russel and Peter Wittenburg, the developers of *ELAN* from Max Planck Institute for Psycholinguistics, who provided us with *ELAN*'s source code and helpful documentation. Also, we would like to thank Laura Buszard-Welcher and Andrea Berez from the Electronic Metastructure for Endangered Languages Data project for their constructive comments on *OntoELAN*.

**Internet Session: LINGUIST List**

LINGUIST List, http://linguistlist.org

The LINGUIST List is dedicated to providing information on language and language analysis, and to providing the discipline of linguistics with the infrastructure necessary to function in the digital world. LINGUIST maintains a website with over 2000 pages and runs a mailing list with over 21,000 subscribers worldwide. LINGUIST also hosts searchable archives of over 100 other linguistic mailing lists and conducts research projects for the development of language engineering tools and recommendations of best practice for digitizing endangered languages data.

***Interaction*:**

Follow the “Language Resources” link and then the “Language Search” link to retrieve all languages spoken in USA. Note that some of the returned languages are marked as “Extinct” and some as “Near Extinction”. Follow the link for one of near extinction languages to see information about that language, explore the language page and check out the links provided for additional information on the language to get a feeling what kind of language data is collected. Note that information on the language is scattered over different linguistic databases and Web portals, and thus requires further integration.

**Internet Session: Electronic Metastructure for Endangered Languages Data (E-MELD)**

E-MELD, http://emeld.org

E-MELD is an NSF-funded project, which aims to create an architecture for digital language archiving, to expedite data access, searching and cross-linguistic comparison. One of the project initiatives is the School of Best Practice, which promotes best practices in digitizing language data. "Best practices" are practices, which are intended to make digital language documentation optimally long-lasting, accessible and re-usable by other linguists and speakers. Recommendations of best practices cover all aspects of digitizing and archiving language documentation, including how to record it, annotate it, catalogue it, store it, and display it in such a way as to respect the intellectual property rights of stakeholders.

***Interaction*:**

Enter the E-MELD School of Best Practice. Explore the "Case Studies" section of the site to learn real examples of preserving language information. Further, explore the "Tool Room" section to learn what softwares are used by linguists for different tasks related to language data (e.g., transcription, video editing and conversion, video alignment). Note that some tools used for similar operations produce outputs in different formats (e.g., plain text, XML, HTML). Therefore, such tools are not interoperable and search over their output data is complicated. Also note that none of the tools is fully suitable for linguists in the sense that it does not support all operations required by linguists.

**Case Study: Video/Audio Annotation with OntoELAN**

In this case study, you are required to annotate a video recording of a non-native English speaker. Your annotation of the recording should contain the monolog transcription, words and their parts of the speech. Start by visiting the OntoELAN homepage (http://database.cs.wayne.edu/proj/ontoelan/), learning instructions on the installation of the tool and reading the "Hands-on Tutorial" section. Deploy and run OntoELAN on your local machine. The workflow of your assignment is described in the following.

- Create an annotation file for one of the sample videos found on the OntoELAN page by selecting the menu "File/New...".
- Create a language profile for your assignment by selecting the menu "File/Profile/New...". Use the GOLD ontology available at http://database.cs.wayne.edu/proj/ontoelan/gold.owl. Add ontological terms that represent parts of speech, such as noun, verb, adverb, adjective and so forth into the language profile. Associate each ontological term with a user-defined term.
- Add three linguistic types by selecting the menu "Edit/Add New Linguistic Type...": TranscriptionType (Stereotype: None), WordsType (Stereotype: Symbolic Subdivision) and GOLDType (Stereotype: Symbolic Association).
- Add three annotation tiers by selecting the menu "Edit/Add New Tier...": Transcription (Parent Tier: None; Linguistic Type: TranscriptionType), Words (Parent Tier: Transcription; Linguistic Type: WordsType) and GOLD (Parent Tier: Words; Linguistic Type: GOLDType; Profile: the path to the language profile).
- Create annotations for the "Transcription" tier: (1) Select the annotation area by clicking the left mouse button, holding it and moving the pointer along the timeline. (2) Right-click the selected area and choose “New Annotation Here” in the pop-up menu. (3) Type the transcription of the recorded talk. (4) Right-click the annotation text and select “Commit Changes”.

- Create annotations for the “Words” tier: (1) Double-click the annotation area next to the “Words” tier. Add the annotation value, the first word of the transcription. (2) Select the slot inherited from the tier's parent by left-clicking on it. Right-click the selected slot and choose “New Annotation After” in the pop-up menu. (3) Add the annotation value, the second word, to the next slot. (4) Continue the creation of new slots and their annotation values until all words of the transcription are added.
- Create annotations for the “GOLD” tier: (1) Double-click the annotation area next to the “GOLD” tier. Annotation slots are inherited from the parent tier. (2) Annotate each slot with the user-defined term (part of speech) from the language profile.
- Save the annotation file by selecting the menu “File/Save”.

***Questions:***

A. What information can one find in an annotation file using OntoELAN search features? How useful is it?

B. How can one implement semi-automatic annotation capabilities in OntoELAN based on one annotation file and based on multiple annotation files available for the same language?

C. How can one annotate a sign language speaker assuming that a sign language ontology and GOLD are available? Describe a workflow.

**Useful Links**

1. OntoELAN homepage, http://database.cs.wayne.edu/proj/ontoelan/
2. LangDL: A Digital Library For Language Engineering And Research, http://database.cs.wayne.edu/proj/langdl/index.html
3. Electronic Metastructure for Endangered Languages Data (E-MELD), http://www.emeld.org
4. LINGUIST List, http://linguistlist.org
5. Rosetta Project, http://www.rosettaproject.org
6. General Ontology for Linguistic Description (GOLD), http://www.emeld.org/gold
7. General Multimedia Ontology, http://database.cs.wayne.edu/proj/ontoelan/multimedia.owl
8. ELAN homepage, http://www.mpi.nl/tools/elan.html
9. W3C Semantic Web, http://www.w3.org/2001/sw/
10. AIS SIGSEMIS: The Semantic Web and Information Systems, http://www.sigsemis.org
11. SemanticWeb.org: The Semantic Web Community Portal, http://semanticweb.org
12. WonderWeb: An Ontology Infrastructure for the Semantic Web, http://wonderweb.semanticweb.org
13. The Semantic Web: A Primer, http://www.xml.com/pub/a/2000/11/01/semanticweb/
14. The Semantic Web: An Introduction, http://infomesh.net/2001/swintro/
15. SchemaWeb: A Directory of RDF Schemas and OWL Ontologies, http://www.schemaweb.info
16. Swoogle: A Semantic Web Search Engine, http://swoogle.umbc.edu

**Further Readings**

**Possible Paper/Essay Titles**

1. Evaluation of existing ontology-based multimedia annotation tools.
2. Ontology-based annotation and semantic search in the multimedia domain.
3. Approaches to automatic and semi-automatic multimedia annotation.
4. Digital libraries: improving recall and precision with ontologies.
5. Knowledge base systems: storing and managing ontologies and annotations.

Figure Captions

*Figure 1*. A snapshot of the OntoELAN main window.

*Figure 2*. OntoELAN Data Flow Diagram.

*Figure 3*. Relationships among some General Multimedia Ontology classes (UML class diagram).

*Figure 4*. An example of the language profile XML document.

*Figure 5*. A snapshot of creating a language profile.

*Figure 6*. A snapshot of annotation tiers in the OntoELAN main window.

*Figure 7*. A snapshot of the tier hierarchy.

*Figure 8*. An example of the XML markup for the OntoELAN annotation.

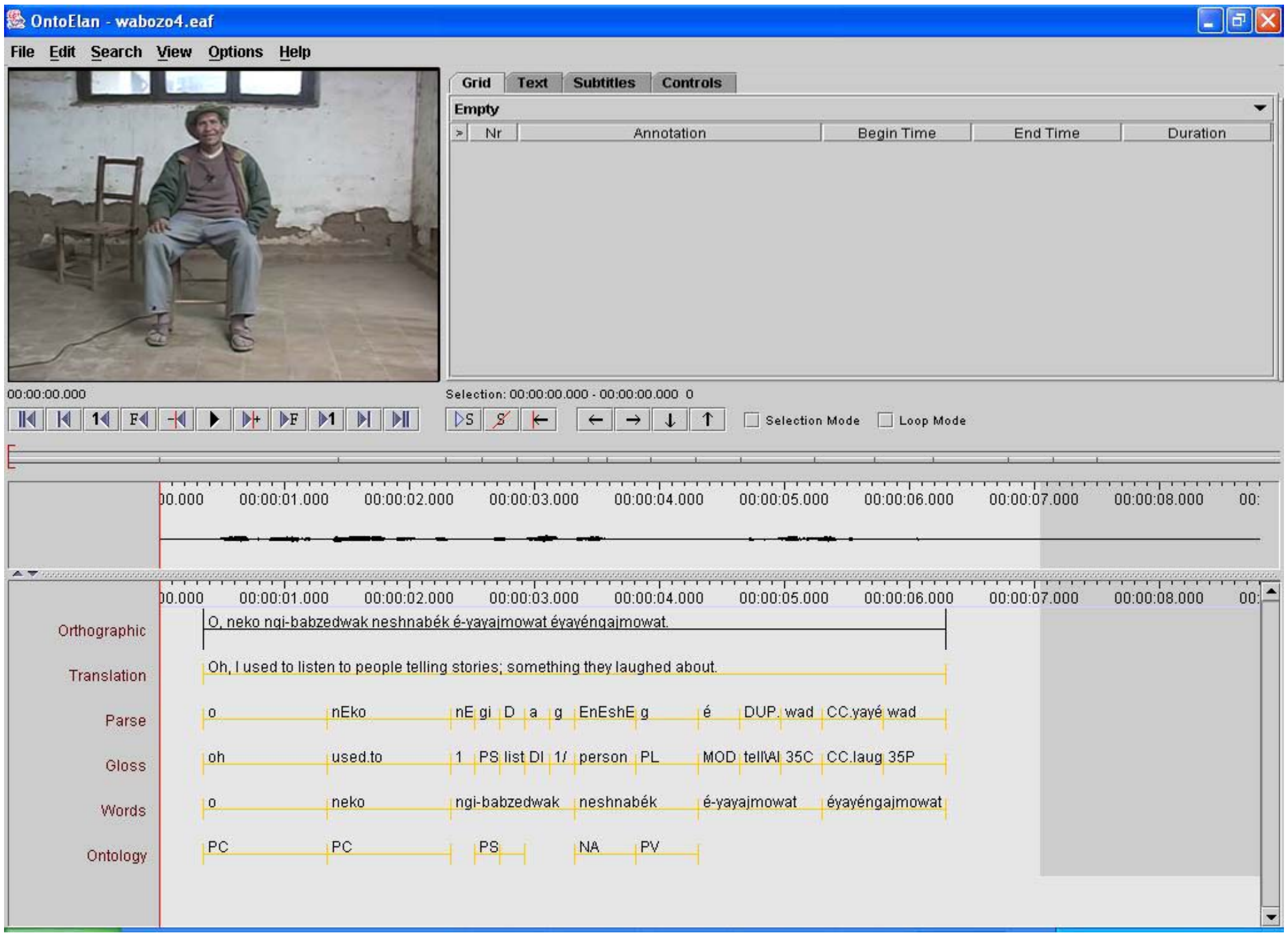

*Figure 1*. A snapshot of the OntoELAN main window

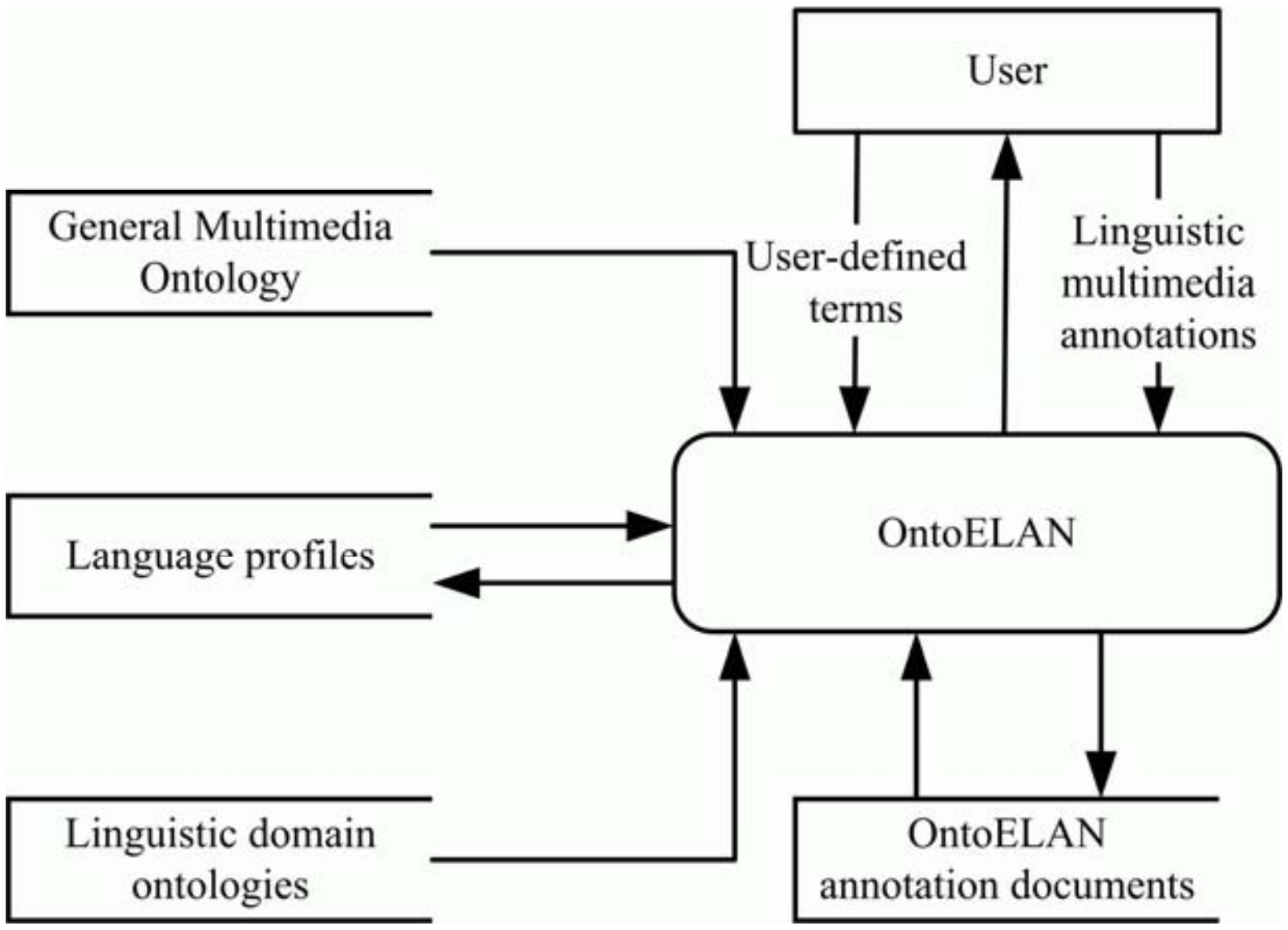

*Figure 2*. OntoELAN Data Flow Diagram

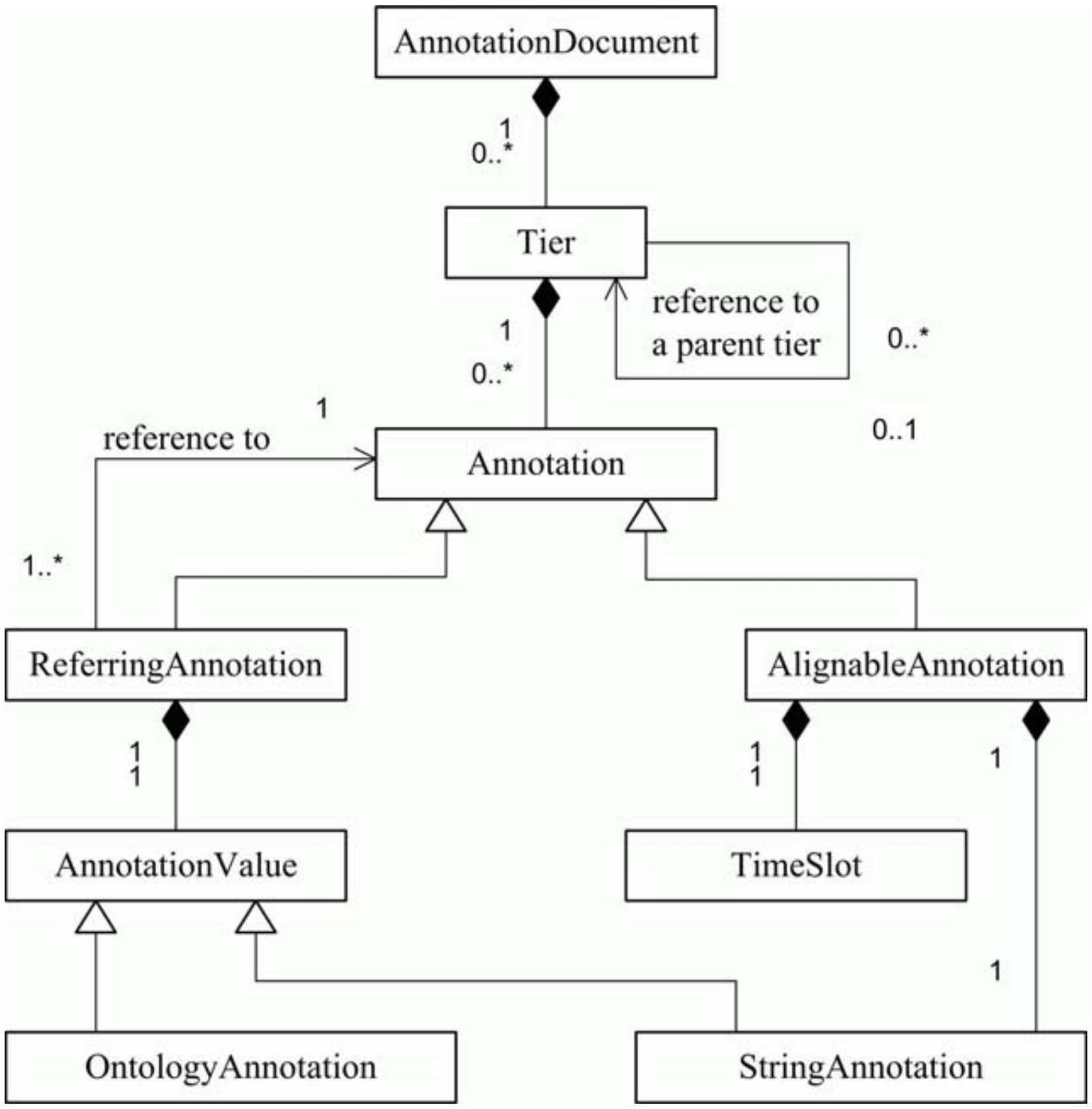

*Figure 3*. Relationships among major General Multimedia Ontology classes (UML class diagram)

```xml
<?xml version="1.0" encoding="UTF-8"?>
<PROFILE AUTHOR="Artem" DESCRIPTION="" VERSION="1.0"
SOURCE= "http://www.u.arizona.edu/~farrar/gold.owl">
<USER_DEFINED_TERM DESCRIPTION="" NAME="NI">
   <ONTOLOGY_TERM NAME="Noun"/>
   <ONTOLOGY_TERM NAME="Inanimate"/>
</USER_DEFINED_TERM>
</PROFILE>
```

*Figure 4.* An example of the language profile XML document

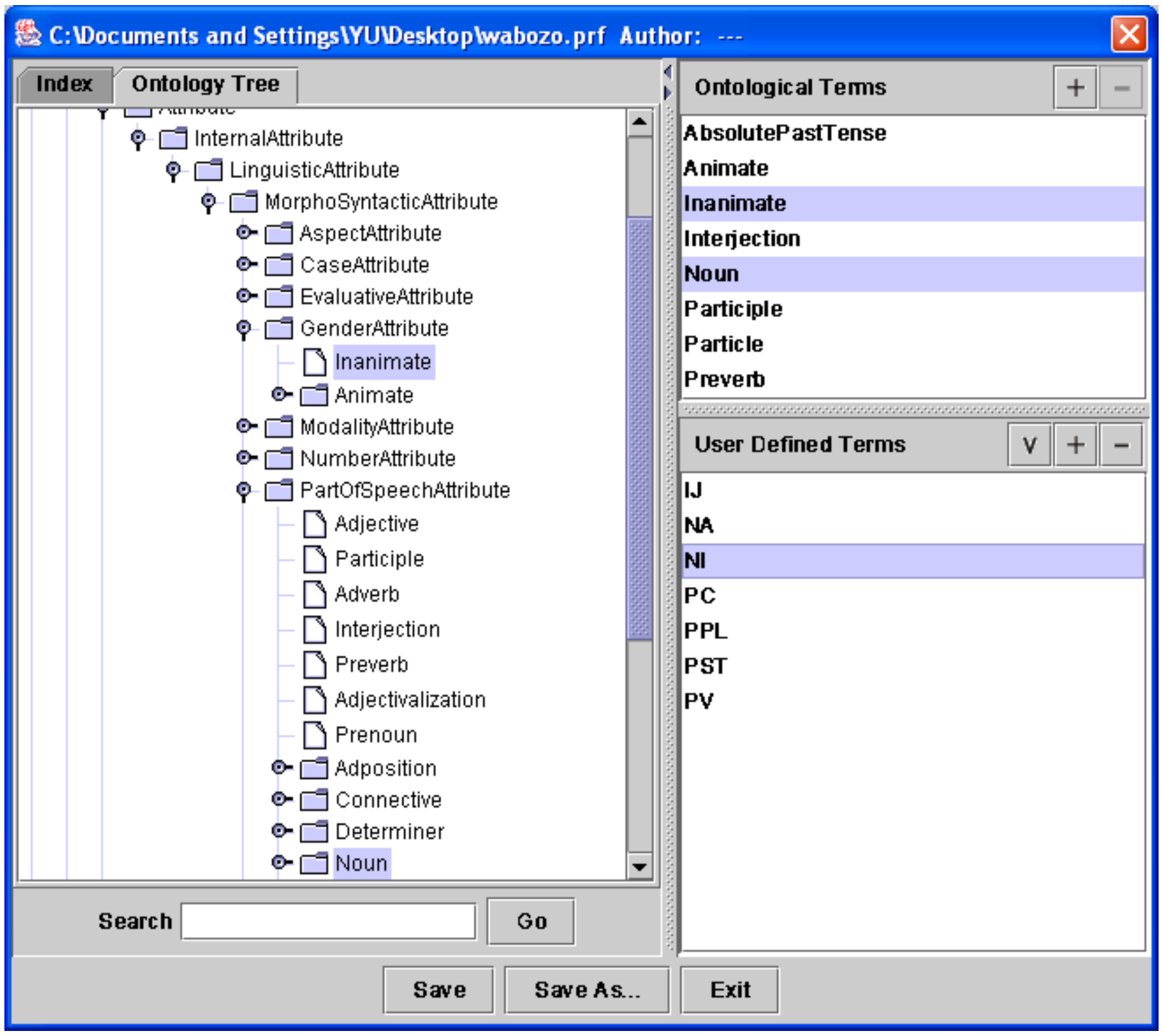

*Figure 5*. A snapshot of language profile creation

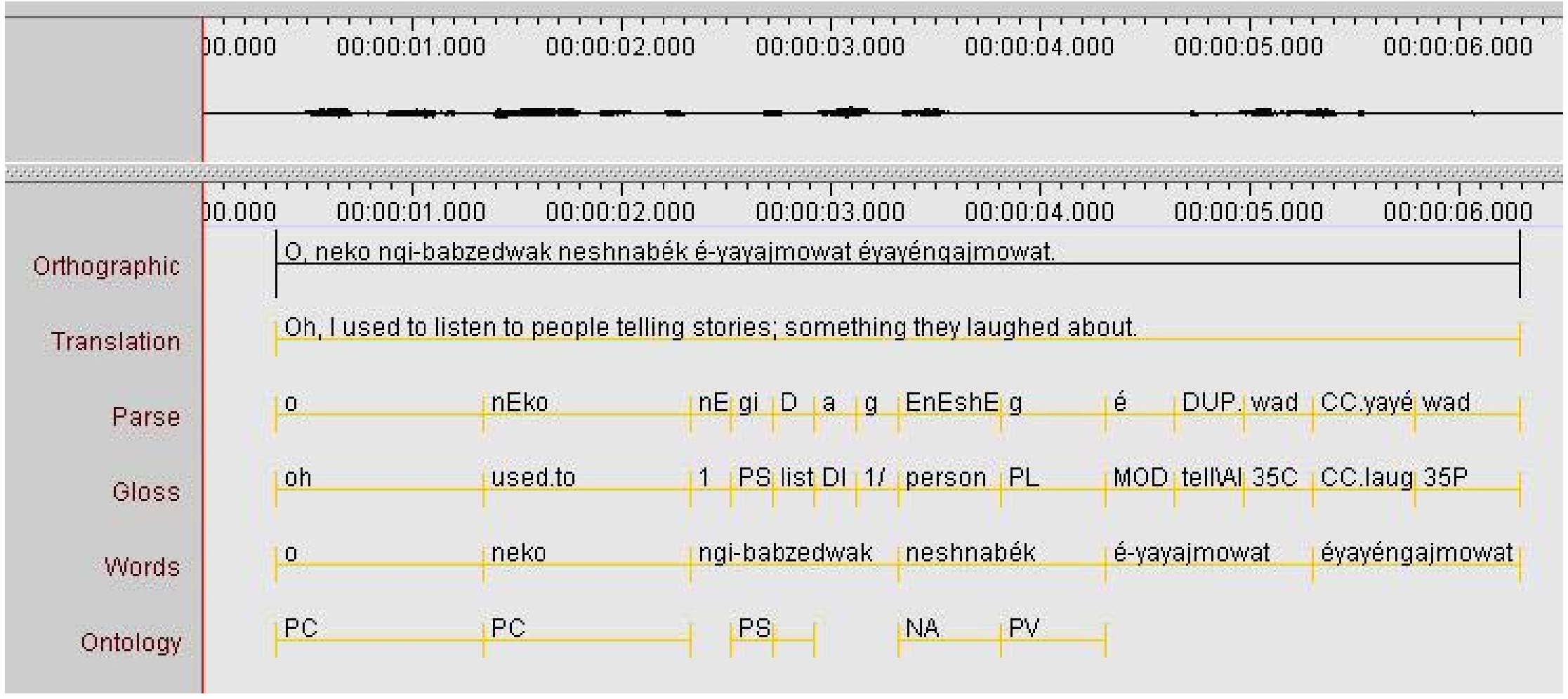

*Figure 6.* A snapshot of annotation tiers in the OntoELAN main window

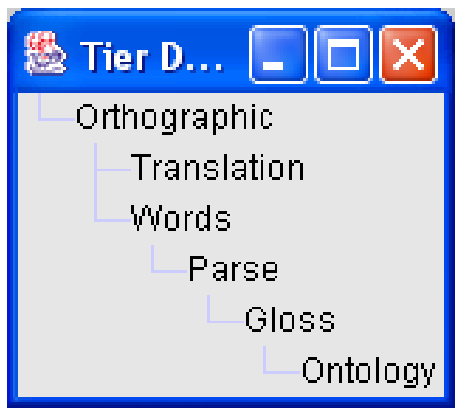

*Figure 7*. A snapshot of the tier hierarchy

```
...
<media:Tier rdf:ID="Ontology">
  <media:hasTierID>Ontology</media:hasTierID>
  <media:hasParent rdf:resource="file:///C:/wabozo4.eaf#Gloss"/>
  <media:hasProfile>C:\wabozo.prf</media:hasProfile>
  <media:hasLinguisticType>
    <media:LinguisticType rdf:ID="ontology">
      <media:hasTimeAlignable>false</media:hasTimeAlignable>
      <media:hasLinguisticTypeID>ontology</media:hasLinguisticTypeID>
      <media:hasConstraint rdf:resource="file:///C:/wabozo4.eaf#Symbolic_Association"/>
      <media:hasGraphicRef>false</media:hasGraphicRef>
    </media:LinguisticType>
  </media:hasLinguisticType>
  ...
</media:Tier>
  ...
<media:RefAnnotation rdf:ID="a42">
  <media:hasAnnotationID>a42</media:hasAnnotationID>
  <media:hasAnnotationRef rdf:resource="file:///C:/wabozo4.eaf#a31"/>
  <media:hasAnnotationValue>
    <media:OntologyAnnotation rdf:ID="a42Value">
      <media:hasUserDefinedTerm>PV</media:hasUserDefinedTerm>
      <media:hasInstances
           rdf:resource="http://www.u.arizona.edu/~farrar/gold.owl#Preverb"/>
      <media:hasOntAnnotationDescription>comments</media:hasOntAnnotationDescription>
      <media:hasOntAnnotationId>e</media:hasOntAnnotationId>
    </media:OntologyAnnotation>
  </media:hasAnnotationValue>
</media:RefAnnotation>
...
```

*Figure 8*. An example of the XML markup for the OntoELAN annotation